\documentclass[fleqn,10pt]{elsarticle}
\title{ Protecting coherence in the non-Hermitian two-level system  }

\author[1]{Wei-Chen Wang}
\author[1,*]{Mao-Fa Fang}
  \address[*]{Corresponding author. E-mail:mffang@hunnu.edu.cn}
 \address[1]{Key Laboratory of Low-Dimensional Quantum Structures and
Quantum Control of Ministry of Education, and Department of Physics,
Hunan Normal University, Changsha 410081, People's Republic of
China}

\begin{document}
\begin{abstract}
\indent
  We have constructed a non-Hermitian two-level system (a $\mathcal{PT}$-symmetric system) in dissipative environments, and investigated the quantum coherence in the non-Hermitian two-level system. Our results show that, quantum coherence can be created by
$\mathcal{PT}$-symmetric systems, even if the initial state of the two-level system is incoherent state. Even though two-level system is interacted with dissipative environments, the quantum coherence exhibits a long-lived revival, and can be protected. We find that the two-level system can obtain more coherence with the coupling strength $\Omega$ increases. And we should point out that the $\mathcal{PT}$-symmetric system can be regarded as a good candidate system for creation of the long-lived quantum coherence in dissipative environments.

\end{abstract}

\flushbottom
\maketitle

\thispagestyle{empty}

\section*{INTRODUCTION}
\setlength{\arraycolsep}{0.1em}
   \indent In general, the Hamiltonian of systems should be Hermitian, if one wants to obtain real eigenvalues of Hamiltonian in quantum physical systems. But in recent years, some authors find that, for the non-Hermitian Hamiltonian, it can still have real eigenvalues if it satisfies $\mathcal{PT}$-symmetry \cite{0013}, while it can have complex eigenvalues when it is in $\mathcal{PT}$-broken phase. The $\mathcal{P}$ stands for a linear parity operator which performs reflection, i.e. $p\rightarrow -p$ and $x\rightarrow -x$. The $\mathcal{T}$ is a linear time-reversal operator has the effect $p\rightarrow -p$, $x\rightarrow x$ and $i \rightarrow -i$ \cite{014}. The non-Hermitian Hamiltonian who satisfies $\mathcal{PT}$-symmetry has some counterintuitive features \cite{015,016}. A non-Hermitian Hamiltonian of two-level system, when a parameter of the system is changed, is able to undergo a transition from a $\mathcal{PT}$-symmetric phase to a  $\mathcal{PT}$-broken phase \cite{017,018}.  For the dynamics of the system, it shows periodic oscillations with time in the $\mathcal{PT}$-symmetric phase region, while it shows exponential growth in the $\mathcal{PT}$-broken phase region. $\mathcal{PT}$-symmetry theory was achieved in many class open system such as optical waveguides \cite{024}, electrical circuits \cite{025}, and laser systems \cite{026}. And recently, $\mathcal{PT}$-symmetry theory was experimentally studied in the quantum system \cite{027}.
   However, there are some controversial results in  $\mathcal{PT}$-symmetric quantum systems, although the $\mathcal{PT}$-symmetric systems open up a new way to process the quantum information.

   Quantum coherence is a new important quantum resource in low-temperature which is studied recently \cite{01,02,03,04,05}. The definition of quantifying coherence in quantum information theories was studied by \cite{06}. Several authors provided many different ways to quantify coherence \cite{07,08,09,023}. In general, coherence and entanglement are fundamental concepts of quantum physics, whose basic idea is superposition principle of quantum mechanics. It is different from entanglement; the coherence is basis-dependent. To protect coherence of a two-level atom, a cavity-based architecture plays a good role \cite{010}. And coherence of two-level systems can be enhanced by interaction with the environments \cite{011}.

    However, almost  these theoretical studies on quantum coherence are based on the Hermitian system, and the quantum coherence in non-Hermitian systems are not examined. At a physical level, $\mathcal{PT}$-symmetric systems are intermediate between closed and open systems. It is worth that we construct a simple non-Hermitian Hamiltonian model and study how it influences quantum coherence.
   \newline \indent Recently the two-level system, which is driven by imaginary field and evolves independently of dissipative environments, was studied by \cite{017,018,012}. In this paper, we consider a two-level system interacting with a quantized radiation field and being driven by imaginary external driving field which is descried by non-Hermitian Hamiltonian $i\Gamma$. And the Hamiltonian of environments and subsystems can be written as:
  \begin{eqnarray}\label{1}
    H&=&H_{0}+i\Gamma ,
    \end{eqnarray}
    where setting $\hbar=1$,
    \begin{eqnarray}\label{02}
    H_{0}&=&\frac{\omega_0}{2}\sigma_z + \sum_{n}^{\infty}a^{\dag}a ,  \nonumber\\
    \Gamma&=&\frac{\Omega}{2}\sigma_{x}\cos\omega_{1} t ,  \nonumber
  \end{eqnarray}
   $\sigma_z$ and $\sigma_x$ are Pauli operator. $a^{\dag}$, $a$ are creation and annihilation operator respectively. $\Omega$ is coupling
strength, and $\omega_1$ is frequency of external filed. So the Hamiltonian of subsystem in rotating-wave approximation can be written as:
  \begin{eqnarray}\label{2}
   H_s&=&\frac{\omega_0}{2}\sigma_z + i\frac{\Omega}{2}(e^{-i\omega_{1} t}\sigma_+ + e^{i\omega_{1} t}\sigma_-).
  \end{eqnarray}
   Where the parameter $\omega_0$ is systemic energy difference, $\sigma_\pm$ denotes the transition operators.

   In next, we will solve the master equation of the system. And we will show that how the quantum coherence evolves in the non-Hermitian two-level system.

\section*{Results and Discussion}
\subsection*{SOLUTION OF  MASTER EQUATION}
\setlength{\arraycolsep}{0.1em}
  For the first, the Schr\"{o}dinger equations for the quantum states $|\Psi\rangle$ and $\langle\Psi|$ with the non-Hamiltonian can be written as:
  \begin{eqnarray}\label{03}
 \frac{\partial}{\partial t}|\Psi\rangle = -iH|\Phi\rangle = -iH_{0}|\Psi\rangle + \Gamma|\Psi\rangle \\
 \frac{\partial}{\partial t}\langle\Psi| = -i\langle \Psi|H^{\dag}=-i\langle \Psi|H_{0} + \Gamma \langle\Psi|
  \end{eqnarray}
  So the dynamics of the density matrix can be recast \cite{013}
  \begin{eqnarray}\label{04}
 \dot{\varrho}(t)=-i[H_{0},\varrho(t)]+\{\Gamma,\varrho(t)\},
  \end{eqnarray}
  where [,] and \{,\} denote the commutator and anticommutator, respectively.
  Thus the master equation of this non-Hermitian open system in interaction picture can be written as:
 \begin{eqnarray}\label{3}
 &\dot{\varrho}_s(t)& = \frac{\Omega}{2}\{\sigma_- + \sigma_+, \varrho_s\} \nonumber\\
 &+& \gamma_{0}(N(\omega)+1)(\sigma_-\varrho_s(t)\sigma_+ - \frac{1}{2}\sigma_+\sigma_-\varrho_s(t)-\frac{1}{2}\varrho_s(t)\sigma_+\sigma_-)\nonumber\\
 &+& \gamma_{0} N(\omega)(\sigma_+\varrho_s(t)\sigma_- - \frac{1}{2}\sigma_-\sigma_+\varrho_s(t) - \frac{1}{2}\varrho_s(t)\sigma_-\sigma_+).
  \end{eqnarray}
  It is worth notice that, when \{,\} is replaced by [,] , the above equation is a master equation of resonance fluorescence \cite{019}.
  On the resonance condition, i.e. $N(\omega)= N(\omega_0)$, where the parameter $N(\omega)$ is the average number of photons in environments with frequency $\omega$.

  We should point out that taking the trace of Eq.(\ref{04}), one can obtain an evolution equation for the trace of $\varrho_s$:
   \begin{eqnarray}\label{05}
   tr\dot{\varrho}_s(t)=2tr\{\Gamma,\varrho_s(t)\}.
  \end{eqnarray}
  Eq.(\ref{05}) shows that the dynamics does not conserve the probability.
  Because the trace of the density matrix $\varrho_s$ is not conserved, we introduce the normalized density matrix as:
   \begin{eqnarray}\label{06}
   \rho(t)=\frac{\varrho_s(t)}{tr\varrho_s(t)}.
  \end{eqnarray}
  It makes sure that the probabilistic interpretation of the density matrix is preserved.
\subsection*{COHERENCE OF THE SYSTEM}
  \setlength{\arraycolsep}{0.1em}
  In order to illuminate how the quantum coherence  evolves in the non-Hermitian two-level system, we introduce two different coherence measures.
  \subsubsection*{Norm of coherence}
  In this subsection, at first, we introduce norm of coherence \cite{06} to measure quantum coherence in the system, which is based on the off-diagonal elements of the desired quantum state, and stands for the fundamental property of quantum interference. For the density operator of an arbitrary quantum state, the formula of norm of coherence is
  \begin{equation}\label{8}
    C(t)=\sum_{i,j(i\neq j)}  |\rho_{ij}(t)|.
  \end{equation}
  Where $\rho_{ij}(t) (i\neq j)$ are the off-diagonal elements of the density operator, and satisfy the quantum physical requirements which conform to coherence measure.

   For the  two-level system initially in the state  $\Psi(0)= \cos(\theta) |1\rangle + \sin(\theta)e^{i\phi} |0\rangle $, the initial norm of coherence is given  by $C(0)=2 \cos(\theta)\sin(\theta)$, (here choose $\phi=0$) from equation (\ref{8}). At first, we consider $\theta=\frac{\pi}{2}$, therefore norm of coherence is zero when $t=0$.  The dynamics of norm of coherence is plotted in Fig.1.
    \begin{figure}
\centering
\begin{minipage}[2a]{0.8\textwidth}
\includegraphics[width=1\textwidth]{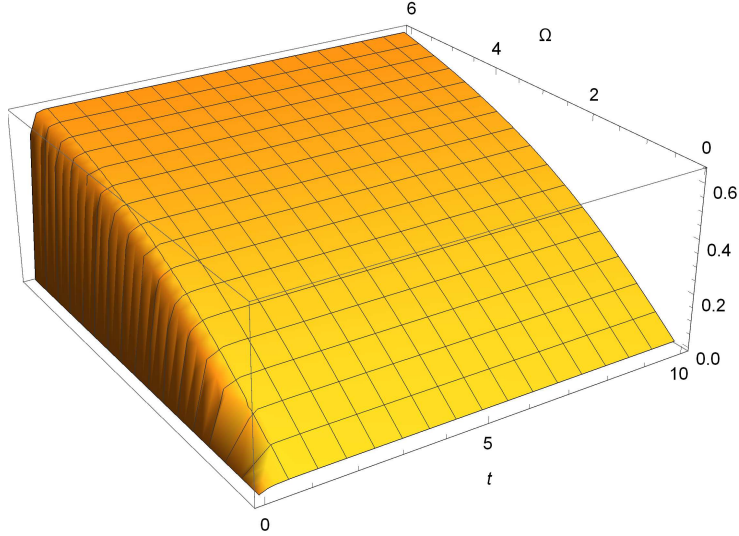} \\
\end{minipage}
\caption{(Color online) The time evolution of norm of coherence of the two-level system. Let us take $N(\omega_0)=5$, $\theta=\frac{\pi}{2}$. Norm of coherence is zero in initial state, and then norm of coherence can be created by external driving field in dissipative environments. The figure also shows that the subsystem is able to obtain more coherence as the coupling strength $\Omega$ increases. }
\end{figure}
    From the Fig.1, it can be seen that, in the initial stage of the time evolution, the norm of coherence is created and then can remain stable at high degree. And the Fig.1 also shows that the subsystem is able to obtain more coherence as the coupling strength $\Omega$ increases.

     Next, let us consider the system in three different situations---no external driving field, external driving field is descried by Hamiltonian $\Gamma$ and external driving field is descried by non-Hermitian Hamiltonian $i\Gamma$. We calculate the master equation (\ref{3}) and display the solution in Fig.2, where we consider $\theta=\frac{\pi}{4}$ in initial state.
    \begin{figure}
\centering
\begin{minipage}[2a]{0.8\textwidth}
\includegraphics[width=1\textwidth]{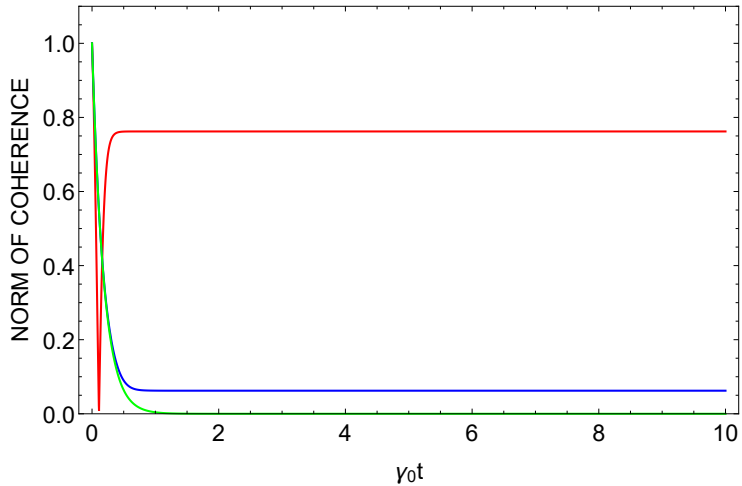} \\
\end{minipage}
\caption{(Color online) The time evolution of norm of coherence of the two-level system. Let us take $N(\omega_0)=5$, $\theta=\frac{\pi}{4}$. $\frac{\Omega}{\gamma_0}=\frac{10}{\sqrt{2}}$. No external driving field (green line), external driving field is descried by Hamiltonian $\Gamma$ (blue line) and external driving field is descried by non-Hermitian Hamiltonian $i\Gamma$ (red line).}
\end{figure}
     From the Fig.2, it can be seen that, in the initial stage of the time evolution, the norm of coherence in three different situations are decaying rapidly, but as time passes, the norm of coherence for imaginary field exhibits a large revival and finally tends to a steady value (red line), while the norm of coherence for real field cannot emerge a revival and eventually trend a minimal value (blue line) and the norm of coherence for no driving field lose all quantum coherence (green line).

 \subsubsection*{The relative entropy of coherence }
 In what follows, we choose another way to quantify coherence of the system which is the relative entropy of coherence \cite{06}. As quantum information entropy is a significant way to describe the quantum information in Hermitian systems, we will use the relative entropy of coherence to interpret quantum features of non-Hermitian systems. Meanwhile the relative entropy of coherence is an important complement to norm of coherence. For the density operator of an arbitrary quantum state, the formula of the relative entropy of coherence is \cite{06}:
\begin{equation}\label{16}
  C_{re}(\rho)=S(\rho_{diag})-S(\rho),
\end{equation}
where $S(\rho)$ is Von Neumann entropy, $\rho=\sum_{i,j}\rho_{i,j}|i\rangle\langle j|$, $\rho_{diag}=\sum_{i}\rho_{i}|i\rangle\langle i|$.

    Even though there are two types of systems---the non-Hermitian system and the Hermitian system, we will show that the formulas of the relative entropy of coherence are same.  A brief proof is following, the  formula of the relative entropy of coherence in the non-Hermitian system can be defined as:
   \begin{equation}\label{18}
    C_{ren}(\rho)= \min_{\delta\in\mathcal{I}} S(\rho\|\delta)=S(\rho_{diag})-S(\rho)+S(\rho_{diag}\|\delta).
   \end{equation}
   Where $\rho$ is a density operator in the non-Hermitian system, $\rho_{diag}$ and $\delta$ are Hermitian operators representing incoherence states, and $\mathcal{I}$ are sets of incoherent quantum states. The minimum of the formula (\ref{18}) can be obtained by taking zero value of third item, and the  formula (\ref{18}) goes back to formula (\ref{16}) formally.
 \begin{figure}
\centering
\begin{minipage}[2a]{0.7\textwidth}
\includegraphics[width=1\textwidth]{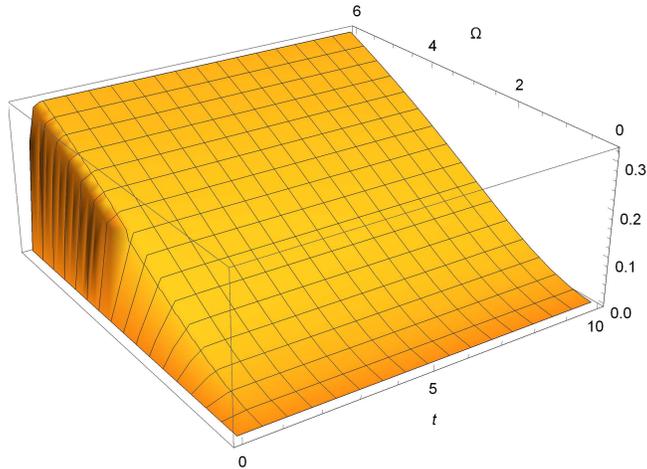} \\
\end{minipage}
\caption{(Color online) The time evolution of norm of coherence of the two-level system. Let us take $N(\omega_0)=5$, $\theta=\frac{\pi}{2}$. Initial state is incoherent state; thus the relative entropy of coherence is zero when $t=0$. And then the relative entropy of coherence can be created by imaginary driving field which is descried by non-Hermitian Hamiltonian $i\Gamma$ in dissipative environments. The figure also shows that the subsystem can obtain more coherence and remain stable with the coupling strength $\Omega$ increases.  }
\end{figure}

      Considering initial state is incoherent state, we can choose $\theta=\frac{\pi}{2}$, and we calculate the master equation and plot the solution in Fig.3. From Fig.3, the relative entropy of coherence can be created by imaginary driving field which is descried by non-Hermitian Hamiltonian $i\Gamma$ in dissipative environments. The Fig.3 also shows that the subsystem can obtain more coherence and remain stable with the coupling strength $\Omega$ increases.
\begin{figure}
\centering
\begin{minipage}[2a]{0.8\textwidth}
\includegraphics[width=1\textwidth]{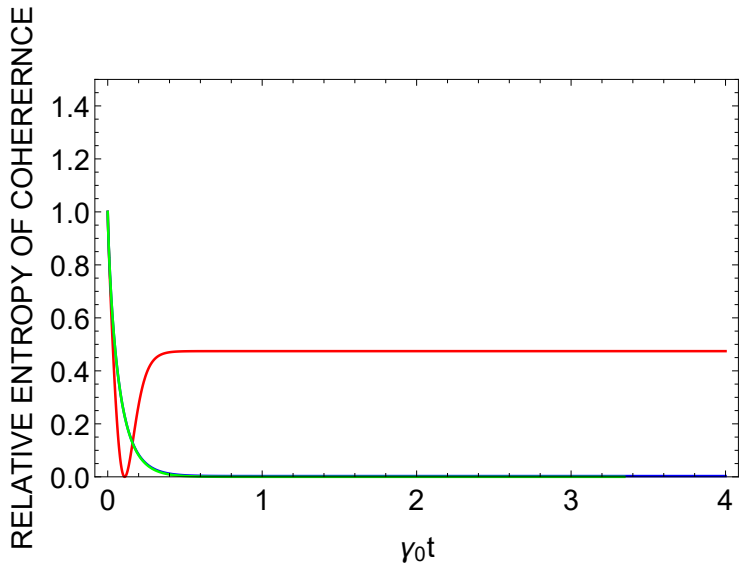} \\
\end{minipage}
\caption{(Color online) The time evolution of norm of coherence of the two-level system. Let us take $N(\omega_0)=5$, $\theta=\frac{\pi}{4}$. $\frac{\Omega}{\gamma_0}=\frac{10}{\sqrt{2}}$. No driving field (green line), real driving field which is descried by Hamiltonian $\Gamma$ (blue line) and imaginary driving field  which is descried by non-Hermitian Hamiltonian $i\Gamma$ (red line) }
\end{figure}

  As same as norm of coherence that we discussed previous, let us induce three different situation---no driving field, real driving field which is descried by Hamiltonian $\Gamma$ and imaginary driving field which is descried by non-Hermitian Hamiltonian $i\Gamma$. Fig.4 presents numerical results for the systemic relative entropy of coherence for no driving field (green line), real driving field (blue line) and imaginary driving field (red line), respectively. In Fig.4, it is evident that, the relative entropy of coherence decays quickly to zero in the all situations. When the driving field is imaginary (red line), it is observed that, the relative entropy of coherence displays a revival and finally tends to a steady value, which as same as the norm of coherence.

\section*{Conclusions}
  \setlength{\arraycolsep}{0.1em}

 In summary, we have investigated norm of coherence and the relative entropy of coherence, and protecting quantum coherence in the non-Hermitian two-level system. We can conclude as follows: First, quantum coherence can be created by $\mathcal{PT}$-symmetric systems, even if the initial state of two-level system is incoherent state. Second, even though subsystem is interacted with dissipative environments, we find that the quantum coherence can exhibit a revival with long time in $\mathcal{PT}$-symmetric systems. Physically, the reason of quantum coherence being created or obtaining revival, is that the general non-Hermitian Hamiltonian can be regarded as an effective tool for the description of source or sink of open systems \cite{013}, thus we deduce that the quantum coherence is probably from another dimensions. Finally, in order to realize this non-Hermitian system in experiment, we can design an experimental facility with the help of the idea which is similar to literature \cite{027}. One can construct an open Hermitian system whose subsystem is described effectively by non-Hermitian Hamiltonian which was discussed in this paper. And then the coherence of quantum system is able to protect in this subsystem. So we should point out that the $\mathcal{PT}$-symmetric systems can be regarded as a good candidate system for creation of the long-lived quantum coherence in dissipative environments.

\section*{References}
\bibliography{sample}

\begin{thebibliography}{1}
\bibitem{0013}    Bender, C. M., and Boettcher, S. Real Spectra in Non-Hermitian Hamiltonians Having $\mathcal{PT}$ Symmetry. Phys. Rev. Lett.\textbf{80} (1998)5243.
\bibitem{014}   Bender, C. M., Brody, D. C. and Jones, H. F. Complex Extension of Quantum Mechanics. Phys. Rev. Lett.\textbf{89} (2002)270401.
\bibitem{015}    Bender, C. M., Brody, D. C., Jones, H. F. and Meister, B. K. Faster than Hermitian Quantum Mechanics. Phys. Rev. Lett.\textbf{98} (2007)040403.
\bibitem{016}    Chen, S. L., Chen, G. Y. and Chen, Y.-n. Increase of entanglement by local PT-symmetric operations. Phys. Rev. A.\textbf{90} (2014)054301.
\bibitem{017}    Gong, J. B. and Wang, Q. h. Stabilizing non-Hermitian systems by periodic driving. Phys. Rev. A.\textbf{91} (2015)042135.
\bibitem{018}    Lee, T. E., Joglekar, Y. N. $\mathcal{PT}$-symmetric Rabi model: Perturbation theory. Phys. Rev. A.\textbf{92} (2015)042103.
\bibitem{024}    Guo, A. et al. Observation of $\mathcal{PT}$-Symmetry Breaking in Complex Optical Potentials. Phys. Rev. Lett.\textbf{103},(2009)093902.
\bibitem{025}   Schindler, J., Li, Ang M., Zheng, C., Ellis, F. M., and Kottos, T. Experimental study of active LRC circuits with  $\mathcal{PT}$ symmetries. Phys. Rev. A.\textbf{84} (2011)040101(R)
\bibitem{026}    Chong, Y. D., Ge, L., and Stone, A. D. $\mathcal{PT}$-symmetry breaking and laser-absorber modes in optical scattering systems. Phys. Rev. Lett. \textbf{106} (2011)093902.
\bibitem{027}    Tang, J. S. et al. Experimental investigation of the no-signalling principle in parity–time symmetric theory using an open quantum system. Nature Photonics doi:10.1038/nphoton.2016.144.
\bibitem{01}    \'{A}berg, J. Catalytic Coherence. Phys. Rev. Lett.\textbf{113} (2014)150402.
\bibitem{02}    Narasimhachar, V. and Gour, G. Low-temperature thermodynamics with quantum coherence. arXiv:1409.7740.
\bibitem{03}    \'{C}wikli\'{n}ski, P., Studzi\'{n}ski, M., Horodecki, M. and Oppenheim, J. Towards fully quantum second laws of thermodynamics: limitations on the evolution of quantum coherences. arXiv:1405.5029
\bibitem{04}    Lostaglio, M., Jennings, D. and Rudolph, T. Description of quantum coherence in thermodynamic processes requires constraints beyond free energy. Nat. Commun.\textbf{6} (2015)6383.
\bibitem{05}   Lostaglio, M., Korzekwa, K., Jennings, D., and Rudolph, T. Quantum Coherence, Time-Translation Symmetry, and Thermodynamics. Phys. Rev. X.\textbf{5} (2015)021001.
\bibitem{06}    Baumgratz, T., Cramer, M., and Plenio, M. B. Quantifying Coherence. Phys. Rev. Lett.\textbf{113} (2014)140401.
\bibitem{07}   Xi, Z. J., Li, Y. M., and Fan, H. Quantum coherence and correlations in quantum system. Scientific reports.\textbf{5} 10922.
\bibitem{08}    Yuan, X., Zhou, H. Y., Cao, Z. and Ma, X. F. Intrinsic randomness as a measure of quantum coherence. Phys. Rev. A.\textbf{92} (2015)022124.
\bibitem{09}    Girolami, D. Observable measure of quantum coherence in finite dimensional systems. arXiv:1403.2446.
\bibitem{023}    Streltsov, A., Singh, U., Dhar, H. S., Bera, M. N., and Adesso, G. Measuring Quantum Coherence with Entanglement. Phys. Rev. Lett.\textbf{115} (2015)020403.
\bibitem{010}    Man, Z. X., Xia, Y. J., and Franco, R. L. Cavity-based architecture to preserve quantum coherence and entanglement. Scientific reports.\textbf{5} 13843.
\bibitem{011}    Ignatyuk, V. V. and Morozov, V. G. Enhancement of coherence in qubits due to interaction with the environment. Phys. Rev. A.\textbf{91} (2015)052102.
\bibitem{012}    Joglekar, Y. N., Marathe, R., Durganandini, P. and Pathak, R. K. $\mathcal{PT}$ spectroscopy of the Rabi problem. Phys. Rev. A.\textbf{90} (2014)040101.
\bibitem{013}    A. Sergi and K. G. Zloshchastiev, non-Hermitian quntum dynamics of a two-level system and models of dissipative environments. Int. J. Mod. Phys. B \textbf{27}, (2013)1350163.
\bibitem{019}    Breuer, H. P., and Petruccione, F. THE THEORY OF OPEN QUANTUM SYSTEMS (OXFORD, London, 2002).
\bibitem{028}    Wehrl, A. General properties of entropy. Rev. Mod. Phys. \textbf{50} (1978)221.
\bibitem{020}    Nielsen, M. A., and Chuang, I. L. Quantum Computation and Quantum Informaton (CAMBRIDGE, London, 2010).


\end{thebibliography}

\bibliographystyle{model1a-num-names}

\section*{Author Contributions}
 W.W. contributed the idea. W.W. performed the calculations. M.F. checked the calculations. W.W.
wrote the main manuscript, M.F. made an improvement. All authors contributed to discussion
and reviewed the manuscript.
\section*{Acknowledgements}

 \setlength{\arraycolsep}{0.1em}
This work was supported by the National Natural Science Foundation of China (Grant Nos.11374096).
\section*{Additional Information}
Competing financial interests: The authors declare no competing financial interests
\end{document}